# A set of semantic data flow diagrams and its security analysis based on ontologies and knowledge graphs

Andrei Brazhuk

Yanka Kupala State University of Grodno (Belarus)

**Abstract**. For a long time threat modeling was treated as a manual, complicated process. However modern agile development methodologies and cloud computing technologies require adding automatic threat modeling approaches. This work considers two challenges: creating a set of machine-readable data flow diagrams that represent real cloud based applications; and usage domain specific knowledge for automatic analysis of the security aspects of such applications.

The set of 180 semantic diagrams (ontologies and knowledge graphs) is created based on cloud configurations (Docker Compose); the set includes a manual taxonomy that allows to define the design and functional aspects of the web based and data processing applications; the set can be used for various research in the threat modeling field.

This work also evaluates how ontologies and knowledge graphs can be used to automatically recognize patterns (mapped to security threats) in diagrams. A pattern represents features of a diagram in form of a request to a knowledge base, what enables its recognition in a semantic representation of a diagram. In an experiment four groups of the patterns are created (web applications, data processing, network, and docker specific), and the diagrams are examined by the patterns. Automatic results, received for the web applications and data processing patterns, are compared with the manual taxonomy in order to study challenges of automatic threat modeling.

## 1 Introduction

Threat modeling is a discipline that deals with deep analysis of a computer system or application design from security perspectives in order to create threat models. A threat model is a list of potential threats for the system used to choose and implement security controls, mitigations or security patterns. For a long time the threat modeling was considered as a manual, iterative, and complicated procedure. This procedure could be done at early stages of system lifecycle (requirements, design). Nowadays, wide-spread get fast (agile) software development methodologies, in which number of software products is dramatically increased and lifecycle stages become extremely short. Also, fast software deployment methods exist, in which architectures of applications are described as declarative configuration files, and applications are run automatically. These changes require adding automatic threat modeling approaches, as well as adaptation the threat modeling for application run-time [Van Landuyt, 2021].

A common approach of the threat modeling supposes the use of Data Flow Diagrams (DFDs). A DFD is a view to a system design/architecture via processes, storages, external entities, and data flows between them. Despite diagrams are considered as a way to informally depict a system, DFDs have been successfully used for formal threat modeling methods done by humans [Kharma, 2023]. However, the question 'Are DFDs useful for automatic analysis?' is still open, as well as the question 'How can DFDs be represented for automatic analysis?'. To learn those challenges the research community requires datasets of diagrams and associated threat models in machine-readable formats. The datasets could also be used for creating, evaluating and proving the semi-automatic (and even automatic) threat modelling methods and techniques. Currently, there is lack of machine-readable datasets of data flow diagrams, used in academia.

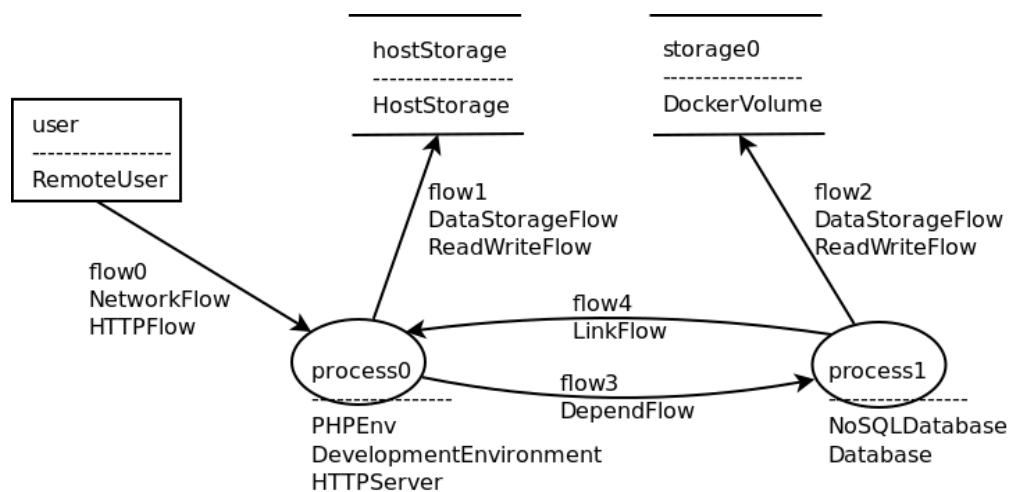

**Figure 1 - An example of semantic data flow diagram**

Inspired by mentioned above questions, this work considers the challenge of creating machine-readable diagrams. Toward the challenge, we have made a set that includes 180 diagrams, called the 'semantic' diagrams because each one contains extra facts about its components and flows. Such a diagram is shown in Figure 1. For example, 'flow0' is 'NetworkFlow' in general, and 'HTTPFlow' in particular; 'process0' is 'HTTPServer', and 'DevelopmentEnvironmet', more precisely 'PHPEnv'.

What can be done with the diagram, shown in Figure 1, from the security analysis perspectives?

In the first case, the multi-level domain specific knowledge can be used for security analysis of flows, processes, storages etc. So, on one level of abstraction, generic network threats of 'process0' can be considered ('flow0' is 'NetworkFlow'); the HTTP specific threats can be relevant on another level ('flow0' is also 'HTTPFlow'). Also, advanced flow paths can be kept in mind, for example, 'process1' may be attacked by 'user' via 'process0' (because of 'flow0' and 'flow3').

In the second case, a concrete data model can be added to the diagram, i.e. some flows can transfer confidential data, some not; some processes work with confidential data, some not. For example, if 'hostStorage' is used to save credentials, 'process0' and 'flow1' can suffer from some specific threats. Also possible indirect actions of 'user' (through 'flow0') and 'process1' (through 'flow4') on 'hostStorage' via 'process0' have to be considered. Note, the domain specific knowledge can be combined with the concrete data model. So, if 'HTTPServer' works with credentials, past specific vulnerabilities of this kind of software can be considered.

One more challenge considered in this work is how the domain specific knowledge can be used to analyse the security aspects of diagrams, i.e. the first case, mentioned above (the second case is left for future research because of dependency from concrete implementations, while we aim to consider common architectures at the moment). Toward this challenge, this work evaluates how ontologies and knowledge graphs can be used to recognize some patterns in diagrams (architectural/security/threat patterns and others – requires extra research to clarify them), and then found patterns can be mapped to security threats. An example of a pattern is HTTP flow between an external entity and a process, e.g. 'flow0' from 'user' to 'process0' in Figure 1. If this pattern is automatically found in the diagram, the system may be vulnerable by the data leak, because HTTP is known as insecure.

Contributions of this work are following:

1) *A set of semantic data flow diagrams that includes 180 items*. Each diagram has three representations: a) YAML file that can be used to draw a picture like Figure 1, or might be converted to another format, b) OWL (Web Ontology Language) ontology that can be used to reason extra facts about the diagram, c) RDF (Resource Description Framework) graph that contains 'full' knowledge about the diagram (so many fact so it can be possible to reason from the ontology) and can be questioned about various aspects of its design and security by SPARQL requests.

The set has been created from real multi-component application configurations: Docker compose configurations (the docker-compose.yml files) have been used. The files have been obtained from public and private repositories, depersonalized to keep the privacy, and converted to three representations by a software tool. The set is freely published (https://github.com/nets4geeks/DockerComposeDataset), and we believe that it can be used for various research in the threat modeling field. Despite the diagrams have been created from particular cloud system configurations, we have tried to make them domain agnostic, and Docker specific aspects are explain in this paper. The diagrams have been manually classified into five categories related to the web applications and data processing. This taxonomy allows to define type of a background application, what gives the first view of the security analysis. So, the diagrams can be considered as a 'wild' set with some ground truth (manual taxonomy).

2) *A test study of usage ontologies and knowledge graphs for security analysis of system configurations*. Four groups of patterns have been created: web applications, data processing, network, and docker specific. A pattern represents features of a diagram (properties of components and relations between them) in form of SPARQL request, what enables its recognition in the semantic representation of diagrams. In the experiment, the RDF graphs of our semantic diagrams' set have been examined by the SPARQL patterns. Automatic results, received for the web applications and data processing patterns, have been compared with the manual taxonomy of the set in order to study challenges of automatic threat modeling.

The structure of the paper is following. Section 2 describes the process of creation of the semantic diagrams' set. Section 3 describes the patterns and the experiment. Section 4 contains evaluation of the automatic threat modeling results. Section 5 describes related work. Section 6 discusses some results of this research.

## 2 A set of semantic data flow diagrams

Both manual and automatic approaches of threat modeling are known in academia and industry. However, there is lack of open data that can be used to create various threat modeling techniques, prove their efficiency, and compare them. In early 2000s Microsoft argued that thousands of DFDs had been creating while working on real software products as part of their Secure Development Lifecycle (SDLC). Obviously, it was a closed dataset that has never been published.

Recent researches operate dozens of diagrams. So, 31 threat models were examined by an academic research [Luburić, 2019] in 2015-2017. A work [Tuma, 2020] has described a full cycle of the threat modeling from creating diagrams to verification threats, also a perfect deep analysis of the automatic and manual aspects, but a resulting publicly available dataset includes 26 annotated design models. A study [Bernsmed, 2022] has been dedicated to the threat modeling in agile teams; the research has involved several organization and universities in Europe in 2017-2020, however most of findings relate to the manual threat modeling.

Our research aims to solve several challenges of forming DFD sets: firstly, how to obtain real data, secondly, how to keep the privacy of commercial sources, and, thirdly, how to increase raw of threat models up to hundreds.

Regarding the first challenge, we use automatic configurations of container based systems as data sources, in particular, the Docker compose configurations. Analysis of system configurations is not new, because several tools are known (see the 'Related work' section). Also, this is some analogy of the threat modeling based on application source code: the automatic threat modeling systems like [Berger, 2019] and [Peldszus, 2022] have lots of raw data in public repositories (github.com, gitlab.com) to work on.

Docker compose is now a part of the Docker project (https://www.docker.com/). It enables a declarative description of several containers used as a multi-container application (like a web application and a background database) in a single file (docker-compose.yml). Using a single command, it can be possible to create and start all the services from the single application configuration. The docker-compose.yml follows the YAML (Yet Another Markup Language) format.

We have used both the github.com and gitlab.com open repositories, as well as several enterprise repositories (about 30% items of the set) in order to collect 180 docker-compose.yml files. To keep the privacy, our set does not include raw docker-compose.yml files, only 'depersonalized' artifacts (data flow diagrams) in several formats (the answer to the second challenge). We have collected 180 configurations, and got an automatic method (with few manual procedures) to produce more diagrams [Ibrahim, 2021] (the solution towards the third challenge).

As we follow an ontological approach of automatic threat modeling [Brazhuk, 2020], [Brazhuk, 2021], each of 180 resulting diagrams is represented as OWL (Web Ontology Language) ontology, what enables automatic reasoning of extra facts, and RDF (Resource Description Framework) graph, what allows the use of knowledge graph based techniques. Either OWL or RDF is a 'semantic' way to deal with data, so the set contains 'semantic' data flow diagrams. DFDs also are described in form of YAML documents that can potentially be used by third-party threat modeling tools.

These 180 diagrams have been classified according five categories (also unclassified items exist). Every category identifies both functional and structural aspects of the diagrams, like a web application and a composite web application (see section 2.5). Such generic taxonomy can be considered as the first step of security analysis. For example, a composite web application requires more attention to the threats of the internal items than a simple web application, and both might be affected by users.

### 2.1 Process overview

Processing of a system configuration represented as a docker-compose.yml file leads to creating three representations of a diagram: OWL ontology, RDF graph, and YAML document (see Figure 2). All the transformations are based on a program model and are done by a Java tool.

Following stages of this automatic process can be emphasized:

1) *Creating a program model from docker-compose.yml*. To do so, it requires some domain specific dictionary/taxonomy to recognize:

- categories of used services (like mysql and postgres are SQL databases in particular and Databases in general),

- types of used data stores (e.g. for mysql the '/var/lib/mysql' folder is a data store, and '/etc/mysql' is a config store),

- kinds of listen network sockets (every recognized network connection to port 3306 is a binary network flow between a mysql client and a mysql server).

A 'services.yml' file contains such a domain taxonomy. Note, creating the domain taxonomy is a manual procedure, and such a taxonomy strongly influences further analysis. The 'services.yml' file has been created as a result of an iterative process while the set has been collected.

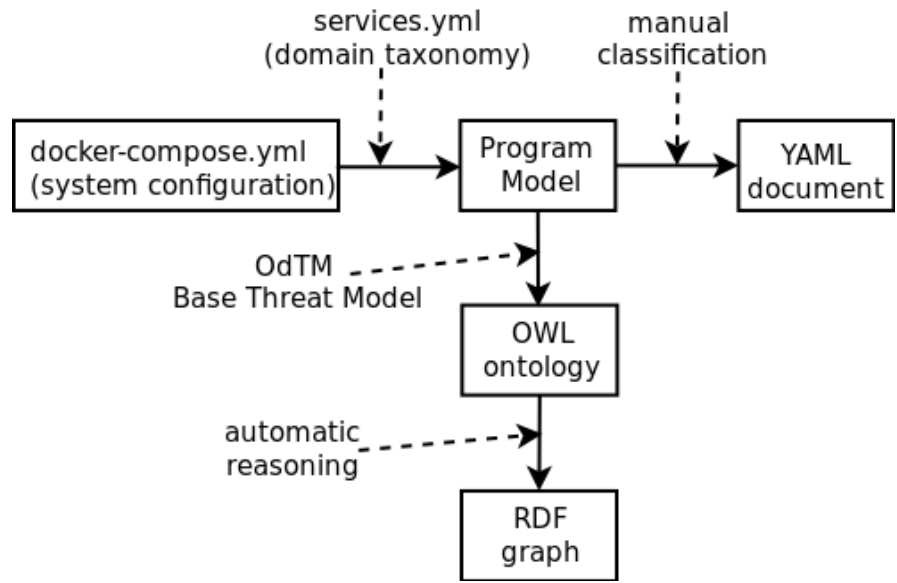

Figure 2 – Process of creating a diagram

2) *Creating OWL ontology from the program model*. The ontological approach supposes a strict formalization of data, what requires a strong model to describe a diagram. The OdTM base threat model is used (OdTM – Ontology driven Threat Modeling Framework) [Brazhuk, 2020]. The base model contains generic concepts and object properties to represent a diagram as an ontology. For example, it can be possible to 'say' that a flow has an external entity as a source and a process as a target.

3) *Creating RDF graph from the OWL ontology*. The OWL ontology contains incomplete knowledge about diagram (like the flow has the process as a target). Automatic reasoning allows to add extra knowledge to the model (like the process is a target of the flow). This extended set of axioms can be saved as an RDF graph. The RDF graph can be asked various questions about its security with SPARQL requests.

The java tool can also create a YAML document from the program model. In this study the YAML documents are used for the manual taxonomy of the diagrams (see below).

**2.2 Creating a program model**

In general to create a program model of a system configuration (docker-compose.yml), it requires:

a) enumerate all the items to build a DFD: processes, data storages, external entities, and flows between them;

b) classify items with a domain specific taxonomy as accurate as possible.

The domain specific taxonomy is placed in the services.yml file, and is used to recognize categorises of services, datastores, and possible network connections. So services.yml define main classes used in further semantic analysis.

In this section we use a simple example to overview the process of creating a diagram for a particular docker compose configuration. A part of used docker-compose.yml file is shown in Figure 3 (Figure 1 depicts this configuration as the diagram modelled with our approach).

**Processes**. The configuration in Figure 3 includes two services (containers): 'web' and 'mongodb' that become processes of the diagram: web is 'process0', and mongodb is 'process1'.

The 'image' property tells what software image is used to create a container (working service). This property can be used to classify a process. So, you can find following definitions in services.yml:

```
services:
 ...
 - name: PHPEnv
   images:
```

```
    - php
  labels:
    - DevelopmentEnvironment
...
  - name: NoSQLDatabase
    images:
      - mongodb
      - mongo
      …
    labels:
- Database
```

'Process0' is based on the 'php' image so it is labeled as 'PHPEnv' (model) and 'DevelopmentEnvironment' (labels) according the current services.yml. 'Process1' uses the 'mongo' image, so it is labelled as 'NoSQLDatabase' and 'Database'.

```
services:
    web:
        image: php:8.0
        volumes:
            - ./app:/var/www/html
        depends_on:
            - mongodb
        ports:
            - 80:80

    mongodb:
        image: mongo:latest
        volumes:
            - dbdata:/data/db
        links:
            - web
```

**Figure 3 – A part of docker-compose.yml**

The first service (web) has the 'ports' property. This property tells that a service is accessible through a network from outside a docker system (a docker host where containers are run). For port '80' the following definition exists in services.yml:

```
ports:
  - name: HTTP
    label: HTTPServer
    value: 80
...
```

From this definition the 'HTTPServer' label can be added to 'process0'. From those we got about 'process0' and 'process1', their YAML definitions can be created:

```
processes:
- name: "process0"
  realName: null
  model: "PHPEnv"
  id: "d3ac6dbf-1836-4df0-9973-07c0c684e677"
  labels:
   - "DevelopmentEnvironment"
   - "HTTPServer"
- name: "process1"
  realName: null
  model: "NoSQLDatabase"
  id: "e2d22f9f-b3ea-456a-a5ad-e06bbedfa44e"
  labels:
   - "Database"
```

**Datastores and data flows**. The 'volumes' property of a service description (see Figure 3) is used to describe storages used by processes. In our model two kinds of storages are considered:

- *Host storage*. A host storage is a folder on a docker host. It is declared as a legacy approach, but it still used because its convenience (you can edit container files from the docker host). Host storages start from dot (.) or slash (/), so the record './app:/var/www/html' (the 'web' service) means the host folder './app' is mapped to the '/var/www/html' folder inside the container. To simplify the analysis, in our model all host folders used by containers are considered as single storage ('hostStorage').

- *Docker volumes*. It is a type of storage implemented by Docker and abstracted for users. Docker volumes can have names or be anonymous. For example, the 'mongodb' container uses the 'dbdata' volume that is connected to the container as '/data/db'.

YAML definitions of storages are:

```
storages:
- name: "hostStorage"
  model: "HostStorage"
  realName: null
  realService: null
  id: "9377dd2b-4a6e-4ba8-8360-6027c8cb2011"
- name: "storage0"
  model: "DockerVolume"
  realName: null
  realService: null
  id: "76c3aa72-5f33-4fe8-83af-f49b4105814b"
```

Also flows between processes and storages can be defined. For example, 'process0' uses 'hostStorage', and this leads to the presence of a new flow ('flow1'). Its YAML definition is:

```
- name: "flow1"
  model: "DataStorageFlow"
  realPortMapping: null
  id: "145f5b24-f7bf-4aa8-a014-dadf3d426f73"
  localPort: null
  source:
    name: "process0"
    id: "d3ac6dbf-1836-4df0-9973-07c0c684e677"
  target:
    name: "hostStorage"
    id: "9377dd2b-4a6e-4ba8-8360-6027c8cb2011"
  labels:
  - "ReadWriteFlow"
```

'Flow1' is classified as 'DataStorageFlow', because services.yml has a section that maps container path (/var/www/data) to this concept:

*datas:*
  *...*
  *- /data/db*
  *...*
  *- /var/www/html*

The services.yml file has also definitions that allow to recognize certificate storages ('CertStorage'), configuration storages ('ConfigStorage'), and log storages ('LogStorage').

By default container folders are connected in read-write mode ('ReadWriteFlow'), however the 'ro' postfix tells to connect external storage in read-only mode ('ReadOnyFlow').

**Network flows**. We emphasize services published by the docker host and accessible from outside (potentially from the Internet), i.e. possible external connections. External connections are paid attention because they are much more insecure; and more secure internal network communications between containers are ignored in our model. Network flows suppose a 'RemoteUser' concept (an external entity) that can be either legitimate user or malicious one. Its definition ('user') in YAML is:

```
externals:
- name: "user"
```

model: "RemoteUser"
id: "4fa525fd-2f97-4710-a2d4-67793bb31f3d"
```

Every network flow goes from the 'user' to a process that has an open port:

```
flows:
- name: "flow0"
  model: "NetworkFlow"
  realPortMapping: null
  id: "0ef26489-8cd6-40b4-9a41-803076ae84bd"
  localPort: null
  source:
    name: "user"
    id: "4fa525fd-2f97-4710-a2d4-67793bb31f3d"
  target:
    name: "process0"
    id: "d3ac6dbf-1836-4df0-9973-07c0c684e677"
  labels:
  - "HTTPFlow"
  realStorageMappings: null
```

For the configuration (Figure 3) this is the port 80, described in services.yml as 'HTTPFlow' (labels). And it is also said that 'flow0' is 'NetworkFlow' (model).

**Dependencies between processes**. Docker compose does not require strict description of dependencies between services. At the same time creating of diagrams is based on precise definition of flows in order to recognize how data move between items of a diagram. So, missed dependencies in the automatic analysis be added by the manual analysis.

Following two properties of a docker compose configuration may indirectly indicate that two services has common data flows:

- The '*depends_on'* property. If a service depends on another service, it means that Docker compose run depended services in order, and depended services will be run before the 'main' service.

- The '*links'* property. It specifies symbolic names by which services are known to each other.

In the configuration (Figure 3), 'process0' depends on 'process1', what creates a flow between them (DependFlow). Its YAML description is:

```
flows:
...
- name: "flow3"
  model: "DependFlow"
  realPortMapping: null
  id: "1eccc998-ef56-4856-b652-d33e8fe397b6"
  localPort: null
  source:
    name: "process0"
    id: "d3ac6dbf-1836-4df0-9973-07c0c684e677"
  target:
    name: "process1"
    id: "e2d22f9f-b3ea-456a-a5ad-e06bbedfa44e"
  labels: null
  realStorageMappings: null
```

And 'process1' has a link to 'process0' (LinkFlow), what in YAML is:

```
flows:
...
- name: "flow4"
  model: "LinkFlow"
  realPortMapping: null
  id: "484b276b-59a0-49b7-b491-1612a03e90af"
  localPort: null
  source:
    name: "process1"
    id: "e2d22f9f-b3ea-456a-a5ad-e06bbedfa44e"
  target:
```

```
  name: "process0"
  id: "d3ac6dbf-1836-4df0-9973-07c0c684e677"
 labels: null
 realStorageMappings: null
```

### 2.3 Creating OWL ontology

The OdTM base threat model [Brazhuk, 2020] as OWL ontology enables semantic interpretation of diagrams. The ontology in the functional syntax has been freely published as the OdTMBaseThreatModel.owl (https://github.com/nets4geeks/OdTM).

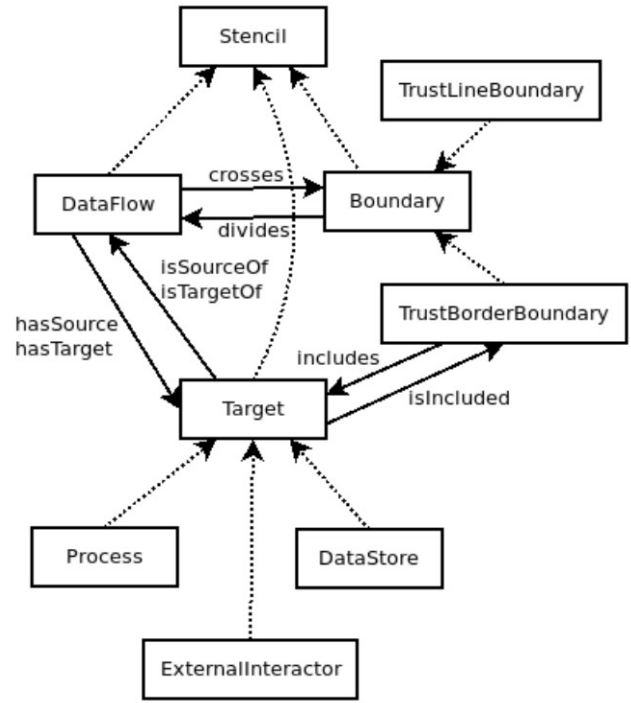

**Figure 4 – Stencils and their properties of OdTM Base threat model**

Figure 4 shows an informal schema of a part of the base threat model responsible for semantic interpretation of diagrams. A common approach (OWASP, Microsoft) tells that a data flow diagram (DFD) consists of 'Stencils'. 'Targets' represent architectural components, connected by flows ('DataFlow'), which might be restricted by 'Trust Boundary'. There are three types of Targets in the model: 'Process', external entity ('ExternalInteractor') and 'DataStore'. The 'DataFlow' concept is defined by its starting and ending nodes (Source and Target). To model the nodes (end points), the 'hasSource' and 'hasTarget' properties are used.

To model a diagram, a new ontology is created, that imports the base threat model. Than all the axioms have to be added:
- instances of processes, storages (hostStorage and volumes), also the 'user' instance as an external entity;
- associations of items to their concepts (model and labels define classes to which an instance belongs to);
- definitions of flows by the 'hasSource' and 'hasTarget' properties; associations them to concepts;
- associations of other local concepts to the concepts of the base model.

For example, the description of the network flow ('flow0') between 'user' and 'process0' with the functional syntax of OWL is like:

```
ClassAssertion(:HTTPFlow :flow0)
ClassAssertion(:NetworkFlow :flow0)
ObjectPropertyAssertion(<http://www.grsu.by/net/OdTMBaseThreatModel#hasSource> :flow0 :user)
ObjectPropertyAssertion(<http://www.grsu.by/net/OdTMBaseThreatModel#hasTarget> :flow0 :process0)
```

And description of 'process0' is like:

```
ClassAssertion(:DevelopmentEnvironment :process0)
ClassAssertion(:HTTPServer :process0)
ClassAssertion(:PHPEnv :process0)
ObjectPropertyAssertion(<http://www.grsu.by/net/OdTMBaseThreatModel#relates> :process0 :hostStorage)
ObjectPropertyAssertion(<http://www.grsu.by/net/OdTMBaseThreatModel#relates> :process0 :process1)
```

The 'relates' object property is quite important here. If two items has a flow, they relate to each other. The property can either be added while flows are generated by Java, or reasoned from the ontology. In this case 'process0' relates to 'hostStorage', because there is the storage flow between them, and 'process0' relates to 'process1' because of the dependent flow.

### 2.4 Reasoning and RDF graphs

The OWL ontology created at the previous stage contains incomplete (explicit) knowledge about a diagram. Automatic reasoning allows to add extra (implicit) knowledge to the model. For example, definition of 'process0' after the automatic reasoning procedure is like:

```
:process0 rdf:type owl:NamedIndividual ,
          bm:AffectedByGenericProcessThreatAsSource ,
          bm:AffectedByGenericProcessThreatAsTarget ,
          bm:Classified ,
          bm:ClassifiedIsEdge ,
          bm:Process ,
          bm:Stencil ,
          bm:Target ,
          :DevelopmentEnvironment ,
          :HTTPServer ,
          :PHPEnv ,
          owl:Thing ;
    bm:isAffectedBy bm:threat_GenericDenialOfService ,
                    bm:threat_GenericElevationOfPrivilege ,
                    bm:threat_GenericInformationDisclosure ,
                    bm:threat_GenericRepudiation ,
                    bm:threat_GenericSpoofing ,
                    bm:threat_GenericTampering ;
    bm:isEdgeOf :flow0 ,
                :flow1 ,
                :flow3 ,
                :flow4 ;
    bm:isSourceOf :flow1 ,
                  :flow3 ;
    bm:isTargetOf :flow0 ,
                  :flow4 ;
    bm:relates :hostStorage ,
               :process1 ,
               :user .
```

Automatic reasoning is based on a set of predefined rules that can be added to an ontology. For example, 'process0' is target of 'flow0', because the properties 'hasTarget' and 'isTargetOf' are defined as inverse in the base model. Also the relation between 'process0' and 'user' has been added because the 'relates' property is symmetric.

The example above shows a peace of the 'full' ontology with the implicit knowledge in form of RDF. The RDF graph can be asked various questions, for example, by SPARQL requests.

### 2.5 The target set and its taxonomy

We have collected 180 docker-compose.yml files (both public and commercial) and processed them by the Java tool. For each configuration we have got a three 'depersonalized' representations of a diagram:
- YAML file (as section 2.2 describes);
- OWL ontology with the explicit knowledge about the diagram (see section 2.3);
- RDF graph both with the explicit and implicit (reasoned) knowledge (see section 2.4).

The set is freely available in the '/clear' folder of the github repository (https://github.com/nets4geeks/DockerComposeDataset).

The java tool, used for creating the set, is published (https://github.com/nets4geeks/OdTM) in the '/applications/parseDockerCompose' folder.

Understanding of what an application does and its structure is the first step of its security analysis. To get some order for the semantic diagrams, an informal criteria (preliminary taxonomy) are added to them. Each criterion informally depicts both functional and structural aspects of a given diagram, and refers to architectural/threat/security patterns that are more precise indicators of security aspects. Section 3 describes an approach of definitions and use the SPARQL-based patterns for the threat modeling.

We use five criteria in the taxonomy:

1) *Web Application*. A single web application based on a database.

2) *Composite Web Application*. A complex application based on a database that might be divided to several components, like frontend and backend. It might be an application that uses extra components to provide advanced services for users, for example instant messaging or project management.

3) *Data collecting*. Represents a set of services used to collect data. May include a service for preparing data (like log parser) and a database to save collected data. It is common to use a document database for saving data (Elasticsearch or Solr).

4) *Data visualizing*. A set of services used to visualize data. It is based on a database where data are, and uses a visualization software to represent data in form pleasant for customers (e.g. Kibana or Grafana).

5) *Complex data processing*. A complex data processing system able to collect, store and visualize data, i.e. some combination of 3) and 4).

Criteria 3), 4), 5) are different from 1) and 2) because of use special software for data processing like Elasticsearch or Kibana, what could have some security consequences.

The YAML files have been manually classified with one or more taxonomy criteria by an external expert in Cloud computing.

The expert followed several rules. Firstly a diagram can fall in two categories, however, if a diagram falls in criterion 2), it does not associated with criterion 1); also, if a diagram falls in criterion 5), it does not associated with criteria 3) or 4). Secondly, some of diagrams was unlabelled because did not fall in any category. Thirdly, the expert used the 'depersonalized' representations, but in several cases he was allowed to view raw configuration in order to better understand 'the nature of a system'.

So, the taxonomy represents the expert (human) view to the diagrams, but this view is primary based on the current services.yml file (i.e. current domain taxonomy created by the author of the paper). The taxonomy can be found on the github repository in the '/labels' folder. And Table 1 shows its statistics.

*Table 1 - Statistics of the set of semantic diagrams*

|   | **Taxonomy criterion** | **Number of diagrams** |
|---|---|---|
| 1 | Web Application | 69 |
| 2 | Composite Web Application | 20 |
| 3 | Data collecting | 5 |
| 4 | Data visualizing | 12 |
| 5 | Complex data processing | 7 |

Table 1 shows that the resulting set includes more descriptions of web based systems, less descriptions of data processing systems, and some unclassified items (about 37 % of the set). So, the set can be considered as a 'wild' set of diagrams that represents various kinds of applications by their functions and structure. Potential use cases of the set can be various informal threat modeling methods, and semi-automatic ones based on domain specific languages, first-order logic, ontologies and graphs.

### 3 Automated threat modeling based on ontologies and knowledge graphs

In this study, the 'pattern' term means features of a diagram (properties of stencils or/and relations between stencils) that can be automatically recognized in a semantic representation of the diagram. A pattern can be associated with a relevant set of threats (see [Brazhuk, 2021]).

Note, the study is focused on patterns that describe system features rather threats associated with those patterns. Collecting and precisely defining threats will be a topic of further research as well as more precise definitions of various patterns.

We have combined two closed approaches to define and recognize patterns in the semantic diagrams:

- OWL and automatic reasoning with background Description Logics (DLs);

- RDF and SPARQL requests.

The Ontology driven threat modeling (OdTM) framework [Brazhuk, 2020] supposes the use of automatic reasoning under domain specific ontological representations of diagrams in order to find relevant patterns. Potential threats have to be associated with some patterns, and the automatic reasoning procedures have to be able to find these patterns in a given ontology (diagram).

For example, lets consider processes as insecure that use the HTTP protocol (because HTTP transfers data in plaintext), and associate them by an appropriate threat. With an ontology editor like Protege (https://protege.stanford.edu/), a pattern (template) for this threat can be defined by Description Logic (DL) as Figure 5 shows:

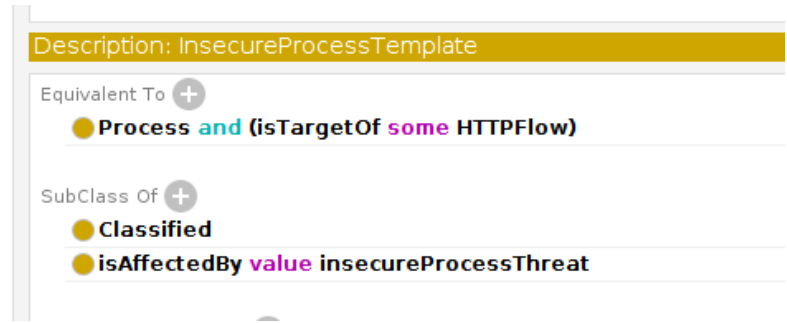

Figure 5 – A simple threat pattern in Protege

The equivalence expression (Figure 5) tells the template to match every instance of the 'Process' concept that is a target of the HTTP flow ('isTargetOf some HTTPFlow'). The subclass expression associates every matched process with the 'insecureProcessThreat' instance by the 'isAffectedBy' property.

A crucial challenge of the automatic reasoning based on Description Logics (DLs) is Open World assumption (OWA) used by DLs based systems (e.g. see work [Cauli, 2021]). In particular, OWA restricts the use of negation in DL expressions, and an expression like shown below 'does not work' as it might be expected:

(isTargetOf some HTTPFlow) and (not (isTargetOf some HTTPSFlow))

The expression above could be used to find stencils that have HTTP flows (insecure) and do not have HTTPS flows (secure). However, a DL reasoner supposes that HTTPS flows exist unknown by the knowledge base (KB); and the reasoner returns nothing because of OWA.

To overcome the OWA challenge, it requires a reasoning system that follows a Closed World assumption (CWA) in some form (i.e. KB is supposed to have all the pieces of knowledge regarding a considered item). A simulation of closed-world reasoning can be used, and even core-closed knowledge bases are under heavy development [Cauli, 2022]. Waiting the valuable progress in this field, we propose to use the RDF representation of ontologies combined with automatic reasoning. If represent all the axioms of an ontology (explicit and implicit) as an RDF graph, it can be possible to ask the knowledge base by the SPARQL queries, what follows the close world assumption and 'works as expected':

```
SELECT ?process
WHERE {
  ?process rdf:type :HTTPServer .
  MINUS { ?process rdf:type :HTTPSServer } .
}
ORDER BY ?process
```

The CWA feature of SPARQL allows precise filtering of results with existing tools, like Apache Jena (https://jena.apache.org/) used in this study.

### 3.1 Groups of patterns

We have created four groups of patterns for the set of semantic diagrams. Every pattern can be automatically recognized by a SPARQL request used to examine RDF graph (full reasoned ontology as section 2.4 describes). The requests use concepts and instances from the OdTM base threat model (the 'b:' prefix), as well as from the services.yml file (the domain taxonomy).

**Docker specific patterns**. Docker and container security is a 'hot' topic both in academia [Minna, 2022] and industry. Several practical guides and checklists exist that consider technological aspects of their security, like OWASP Docker Top 10 (hhttps://github.com/OWASP/Docker-Security) and CIS Docker Benchmark Profile (https://dev-sec.io/baselines/docker/). However, our work is focused on common security aspects (application architectures), so we have added a few specific aspects of the Docker security in order to show the possibility to learn the specific aspects. More specific patterns might require the extension of the ontology creation process in order to keep extra configuration features.

*(1-1) Read-only access to Docker socket and (1-2) Read-write access to Docker socket*. Docker socket is a communication path between administrative client tools and the Docker service (i.e. process that manages containers on the host). Mounting the Docker socket as a storage inside a container is considered as security violation in the literature [Rice, 2020]. The program model is able to describe the use of Docker socket and SPARQL can be used to find the patterns. For example, to recognize the flows that satisfy the pattern (1-2), the query can be like:

```
SELECT ?source ?flow ?target
WHERE {
  ?flow b:hasTarget :dockerSocket ;
      rdf:type :ReadWriteFlow .
  ?source b:isSourceOf ?flow .
  ?target b:isTargetOf ?flow .
}
ORDER BY ?source
```

*(1-3) Read-write configuration storage*. In general, no need to change the configuration data by a container, so the configuration volumes used in the read-write mode might be considered as some security issue. To find this patterns, SPARQL can be used:

```
SELECT ?source ?flow ?target
WHERE {
  ?flow rdf:type :ConfigStorageFlow ;
      rdf:type :ReadWriteFlow .
  ?source b:isSourceOf ?flow .
  ?target b:isTargetOf ?flow .
}
ORDER BY ?source
```

**Network patterns**. There are some primitives that relate to the network aspects of security. These patterns come from open ports of services, accessible from outside of the Docker host. They are modelled as network flows from the user (an external entity) to the processes.

*(2-1) HTTP flow*. It is known that the use of HTTP is insecure. We have found that real system configurations still use HTTP. However, to decrease number of false positives, the cases have to be excluded where HTTP is used for compatibility and the initial HTTP request is forced to the HTTPS port. The following query can be used:

```
SELECT ?source ?flow ?target
WHERE {
  ?flow rdf:type :HTTPFlow ;
      b:hasTarget ?target ;
      b:hasSource ?source .
  MINUS {
    ?flow1 rdf:type :HTTPSFlow .
    ?target b:isTargetOf ?flow1
  }
}
ORDER BY ?flow
```

*(2-2) Database network flow*. Direct access from outside to databases is rarely used, and in many cases this can be considered as some misconfiguration. The request can be used to recognize the outside access to a database:

```
SELECT ?source ?flow ?target
```

```
WHERE {
  ?flow rdf:type :DBFlow ;
      b:hasTarget ?target ;
      b:hasSource ?source .
}
ORDER BY ?flow
```

**Web architecture patterns**. This group aims to find common items of web architecture and their combinations. The 'relates' object property is used to discovery relationship between components, what allows recognition of advance sequences of components that can be affected by complex threats. The web application security is a well-researched topic, and lots of threat and security models exist there [Shahid, 2022], so every pattern in this group can be associated with a relevant list of threats.

*(3-1) User-Websever*. This simple pattern represents interactions between users and a webserver, what is considered as the initial step of most web attacks. To find this patterns, SPARQL can be used:

```
SELECT ?target
WHERE {
  ?target rdf:type :WebServer ;
      b:relates :user .
}
ORDER BY ?target
```

*(3-2) WebServer-Database and (3-3) DevelopmentEnvironment-Database*. Any kind of database (relational, non-relational or document based) can be a background of a web application. In our model, these are either relations between the 'WebServer' and 'Database' concepts or between 'DevelopmentEnivironment' and 'Database'. Choice of the concept depends on the initial software image: 'WebServer' means that the initial image is 'nginx' or 'apache', and 'DevelopmentEnvironment' relates to images that are based on the programming language environments (PHP, Python etc.). Commonly, these types are similar, because a nginx based images has to include some programming language, as well as a PHP based image has to include some web server.

For example, the query of the (3-2) pattern is:

```
SELECT ?target ?target1
WHERE {
  ?target rdf:type :WebServer .
  ?target1 rdf:type :Database ;
      b:relates ?target .
}
ORDER BY ?target
```

*(3-4) WebServer-DevelopmentEnvironment-Database*. This pattern recognizes advanced configurations based on a sequence of three services. It is supposed that a web server holds static data and works as the frontend of a development environment, while the development environment is in charge of dynamic content an interacts with a database.

```
SELECT ?target ?target1 ?target2
WHERE {
  ?target rdf:type :WebServer .
  ?target1 rdf:type :DevelopmentEnvironment ;
      b:relates ?target .
  ?target2 rdf:type :Database ;
      b:relates ?target1 .
}
ORDER BY ?target
```

*(3-5) CacheDatabase-Process and (3-6) WebProxy-Process*. The first pattern relates to a common item (cache) used to make access to content faster by placing data in memory, what might have security consequences. The second pattern relates to configuration where extra component is added between the user and applications for security reasons or to enable load balancing.

The (3-5) SPARQL is:

```
SELECT ?target ?target1
WHERE {
  ?target rdf:type :CacheDatabase ;
          b:relates ?target1 .
  ?target1 rdf:type b:Process .
}
ORDER BY ?target
```

**Data processing patterns**. This group includes the patterns based on services that are used for data management and manipulations, what seems to be a perspective direction of security research [Kilic, 2016] [Abbass, 2019]. Note, SPARQL requests for this group you can find in the set repository in the 'sparql' folder.

*(4-1) DataCollector-Database*. Supposes that a diagram has a component to collect data ('DataCollecter') and a database.

*(4-2) DataVisualizer-Database*. Includes the 'Database' item and a service recognized as 'DataVisualizer' (like Kibana).

*(4-3) DataCollector-Database-DataVisualizer*. Represents an advanced sequence of services that are able to provide advanced data processing.

*(4-4) Database-DatabaseManagement*. Filters cases of the database management system use. 'DatabaseManagement' refers to applications used to direct access of data, what seems to be an issue for productive environments.

*(4-5) MessageBroker-Process*. Recognizes the use of the 'MessageBroker' instance.

### 3.2 The experiment and results

The semantic diagrams (their RDF representations) of the set (see section 2) have been examined by all the described patterns (i.e. by SPARQL requests that satisfy the patterns).

Results are shown in Table 2. The 'Number of diagrams' field shows the number of diagrams where a pattern has been found.

*Table 2 - Experiment results*

|     | Pattern | Number of diagrams |
|-----|---------|--------------------|
| **Docker specific patterns** | | |
| 1-1 | Read-only access to Docker socket | 12 |
| 1-2 | Read-write access to Docker socket | 12 |
| 1-3 | Read-write configuration storage | 52 |
| **Network patterns** | | |
| 2-1 | HTTP flow | 122 |
| 2-2 | Database network flow | 56 |
| **Web architecture patterns** | | |
| 3-1 | User-Websever | 49 |
| 3-2 | WebServer-Database | 24 |
| 3-3 | DevelopmentEnvironment-Database | 40 |
| 3-4 | WebServer-DevelopmentEnvironment-Database | 10 |
| 3-5 | CacheDatabase-Process | 30 |
| 3-6 | WebProxy-Process | 13 |
| **Data processing patterns** | | |
| 4-1 | DataCollector-Database | 5 |
| 4-2 | DataVisualizer-Database | 11 |
| 4-3 | DataCollector-Database-DataVisualizer | 5 |
| 4-4 | Database-DatabaseManagement | 14 |
| 4-5 | MessageBroker-Process | 15 |

## 4 Discussion

Some results of the experiment (see Table 2) can be evaluated heuristically. So, unfortunately many container applications (122 in our set) still use HTTP, and a few of them (12) require the read-write access to Docker socket etc.

Some results from the web architecture patterns and the data processing patterns can be compared with the manual taxonomy of the semantic diagrams' set (see section 2.5), what gives evaluation how effective are automatic patterns in the recognition of common application types. Table 3 shows such the evaluation.

*Table 3 – Recognition of diagram types by semantic patterns*

| | Criteria | Patterns | Total criteria | Detected patterns | Positive | False Positive | False Negative | Precision | Recall |
|---|---|---|---|---|---|---|---|---|---|
| | (1) Web Application +(2) Composite Web Application | (3-2) WebServer-Database (3-3) DevelopmentEnvironment-Database | 89 | 63 | 61 | 2 | 28 | 0.97 | 0.69 |
| | (2) Composite Web Application | (3-4) WebServer-DevelopmentEnvironment-Database | 20 | 10 | 10 | 0 | 10 | 1.00 | 0.50 |
| | (3) Data collecting +(5) Complex data processing | (4-1) DataCollector-Database | 12 | 5 | 5 | 0 | 7 | 1.00 | 0.42 |
| | (4) Data visualizing +(5) Complex data processing | (4-2) DataVisualizer-Database | 19 | 11 | 11 | 0 | 8 | 1.00 | 0.58 |
| | (5) Complex data processing | (4-3) DataCollector-Database-DataVisualizer | 7 | 5 | 5 | 0 | 2 | 1.00 | 0.71 |

Composite web applications (2) can be recognized by the 'WebServer-DevelopmentEnvironment-Database' (3-4) pattern. Web applications (1) can be recognized either the 'WebServer-Database' (3-2) or 'DevelopmentEnvironment-Database' (3-3) patterns.

The 'Web application' diagrams includes the 'Composite web applications' diagrams from the pattern view, because latter includes simpler patterns (e.g. 'DevelopmentEnvironment-Database' is a part of 'WebServer-DevelopmentEnvironment-Database'). The similar reason is to consider 'Complex data processing' (5) as a part of 'Data collecting' (3) and 'Data visualizing' (4) while mapping them to the patterns.

Note, 'Total criteria' means number of diagrams in the group(s) of manual taxonomy, 'Detected patterns' - number of diagrams where a pattern (patterns for '1') has been detected, 'Positive' - number of right results.

The precision is considered as:

'Precision' = 'Positive' / 'Detected patterns'

Hundred percent of precision is a common result for deterministic approaches, like one used in this work. If based on well-formed domain taxonomy, SPARQL requests work as filters and have few false positive result.

The recall is considered as:

'Recall' = 'Positive' / 'Total criteria'

As Table 3 shows, false negatives (not detected criteria) happen. Two reasons can be why the filters miss right results:
- Undetected relations between stencils. A Docker compose configuration does not require specify all the dependencies between services, so 'DependFlow' and 'LinkFlow' are not enough to determine the dependencies. However, the architectural patterns, used in the experiment, relay on relations between components. Solutions might be manually added flows; also advanced graph based techniques that 'predict' relations can be used [Rossi, 2022], [Baghershahi, 2023].

- Narrow domain taxonomy. Learning low recall value for composite web applications, we have found, that the human expert has much more understanding regarding components, than the domain taxonomy has knowledge about them. Add adding extra knowledge the taxonomy significantly decreases number of false negatives.

## 5 Related work

Currently, the threat modeling is considered as semi-automatic process, added primarily at the requirements and design stages of a system lifecycle [Konev, 2022]. Well-known manual methodologies [Kim, 2022], [Al-Momani, 2022], tend to add more formalization to this process.

A recent research [Rodrigues, 2023] offers a privacy threat modeling language in order to make bridge between functionality of domain specific applications and software design. Work [Rao, 2023] offers a threat modeling framework for mobile communication systems. Research [Von Der Assen, 2022] considers challenges of collaborative threat modeling. Work [Yaqub, 2022] offers a STRIDE based approach to the threat modeling of IoT based systems.

The automation of the threat modeling is still a research challenge; the efforts are based on graph theory, Domain Specific Languages (DSL) and rule based languages, First Order Logics (FOL) and Prolog, also Ontologies.

Work [Al Ghazo, 2019] has proposed an algorithm of generation of attack graphs in order to enumerate all sequences in which weaknesses can be used to compromise system security. Research [Alshareef, 2021] has described an implementation of privacy threat modeling based on the graph homomorphism.

Work [Hacks, 2022] describes a meta language based on attack graphs for creation of domain specific languages used for threat modeling. Research [Alshareef, 2022] defines a formal framework for annotating DFDs with purpose labels and privacy signatures; the framework includes a domain specific language to specify the signatures, and an algorithm to check and reason the purpose labels from the signatures.

Work [Seifermann, 2022] presents an extended DFD syntax to model both information flows and access control aspects; the semantic of the syntax is defined by clauses in FOL. Research [Rouland, 2021] has described formalization of STRIDE threat categories using first order and modal logics; they have used UML and a component-port-connector model for system representation.

The ontological approach of the threat modeling seems to be simpler than FOL and Prolog, follows the object paradigm and can be easily extended (OWL, SPARQL, SWRL etc.). The idea of the ontological approach are described in works [Venkata, 2018], [Williams, 2022]. Works [Brazhuk, 2020], [Brazhuk, 2021] have offered an ontology driven framework of the DFD based threat modeling and its use case for cloud systems. An ontology based multi-layer framework for analysis, threat intelligence and validation of security policies in computer systems has been described in work [Vassilev, 2021]. Recent researches in the ontology based threat modeling are dedicated to the automation of security analysis of ICT Infrastructures [Rosa, 2022] and Amazon Cloud Infrastructure [Cauli, 2021].

Work [Engelberg, 2022] proposes a risk propagation approach based on an ontology, semantic representations of intelligent systems, and a method of calculation the risk propagation within a given system. Research [Oliveira, 2022] investigates the capabilities of ontological constructs in enterprise security modeling. Work [Shaked, 2022] describes a method of automated cyber risk identification based on ontologies.

Adding Machine Learning (ML) and Neural Networks (NNs) to the threat modeling is a great open challenge. Despite these technologies are not used in the particular field, security science knows plenty examples of their use [Demirol, 2022]. A recent work [Shashkov, 2023] researches artificial intelligence and machine learning methods for simulation both agents launching attacks and agents using detection and mitigation mechanisms against threats. Research [Chakir, 2023] considers how traditional ML can be added to web based attacks detection. Also, forensic investigations [Touloumis, 2022] and provenance graph learning [Chen, 2022] are in focus. Work [Sworna, 2023] proposes a NNs based framework for security tool API recommendations.

Wide spread of cloud computing has been changing the principles of risk assessment, security design, and operation security of computer systems. Cloud systems require both more precise resource management [Wang, 2022] and advanced cyber threat intelligence [Ananthapadmanabhan, 2022]. An actual direction is creation of a common format for declarative description of cloud applications [Zalila, 2022], [Challita, 2021] in order to unify the security aspects of cloud deployments and add new security methods [Kunz, 2023].

Threat modeling and security aspects of containers are under consideration in the academia realm [Wong, 2023], [Haque, 2020]; this includes wide range of topics. So, work [Alyas, 2022] proposes a Docker

security mechanism to evaluate run-time configuration of containers, as well as scanning of images for vulnerabilities. Research [Leahy, 2022] offers a Zero Trust Architecture variant for secure Docker deployments. Work [Minna, 2022] proposes a a sort of knowledge approach to model cloud deployment security objects and vulnerabilities based on AND/OR logics and graphs. Research [Sergeev, 2022] is dedicated to stress testing of Docker containers.

Work [Ibrahim, 2021] has done a study of more 4000 Docker compose based applications in order to learn how Docker is used in real projects and their architectural features. Research [Zheng, 2021] has offered a method of detecting violations in Docker images. And work [Zhu, 2021] has proposed a Docker security profile generator.

Several works research aspects of Kubernetes security [Shamim, 2020]. Work [Blaise, 2022] proposes a methodology for extracting and evaluating the kubernetes security from a topological graph where nodes and edges are security features; also they have used the ATT&CK framework for risk assessment. Research [Haque, 2022] offers a knowledge based approach to finding misconfigurations in several cloud based platform (Kubernetes, Azure, VMWare, Docker). And work [Budigiri, 2021] investigates the security aspects of network policies in Kubernetes.

Recent growth of the cloud computing technologies and fast software deployment methods creates new challenges of the threat modeling. Automatic methods able to work in run-time should be used [Van Landuyt D, 2021]. Industry proposes several tools based on existing scientific findings towards this challenge like CAIRIS [Faily, 2020], and securiCAD [Wideł, 2022], as well as engineering tools like IriusRisk (https://www.iriusrisk.com/) and ThreatModeller (https://threatmodeler.com/). The automation of the secure development process requires advanced knowledge management approaches, in particular, maintaining and updating open security knowledge bases (attacks, weaknesses, vulnerabilities etc.).

## 6 Conclusions

This work has considered two challenges: firstly, creating the set of semantic data flow diagrams based on real cloud applications and potentially used for evaluation various threat modeling techniques; secondly, usage domain specific knowledge for automatic analysis of the security aspects of such applications.

The diagrams' set can be quite useful to learn deterministic threat modeling methods, however number of items and their classification might be not enough for ML based experiments. Note, number of items can be extended, because the creating diagrams is a well automated process.

The automated threat modeling based on ontologies and knowledge graphs has been tested for a generic use case. For a particular case, it needs to consider real data flows and classify security degree of data, what requires more models and assumptions.

A particular use case (like a technique of adding a real data model), might be considered in further research. Also, the semantic approach strongly requires well formed domain specific threat models, what can be considered as a knowledge management challenge. The third direction of research might be clarification of the 'pattern' concept, because design patterns, security patterns, threat patterns are known as manual entities, their automation is a research challenge.